\documentclass[9pt,conference]{IEEEtran}

\usepackage[preprint]{waspaa25}

\usepackage{bm} 

\usepackage{amsmath,graphicx}
\usepackage{balance}

\usepackage{amsfonts}
\usepackage{algorithm2e}
\usepackage{array}
\usepackage[caption=false,font=normalsize,labelfont=sf,textfont=sf]{subfig}
\usepackage{textcomp}
\usepackage{stfloats}
\usepackage{verbatim}
\usepackage{graphicx}
\usepackage{multirow}

\usepackage[all=normal,mathspacing]{savetrees} 

\newcommand{\1}{\mathbf{1}}

\DeclareMathOperator*{\argmax}{argmax}
\DeclareMathOperator*{\argmin}{argmin}

\newcommand{\x}{\mathbf{x}}

\newcommand{\y}{\mathbf{y}}

\newcommand{\T}{\mathbf{T}}

\newcommand{\V}{\mathbf{V}}

\newcommand{\pbf}{\mathbf{p}}

\newcommand{\Real}{\mathbb{R}}

\newcommand{\m}{\mathbf{m}}

\newcommand{\rbf}{\mathbf{r}}

\newcommand{\taubf}{\boldsymbol{\tau}}

\newcommand{\ji}{\jmath}

\DeclareMathOperator{\EX}{\mathbb{E}}
\usepackage{tikz,pgfplots}
\usetikzlibrary{decorations.pathreplacing,plotmarks}
\pgfplotsset{compat=1.7}
\usepackage{textcomp}
\usetikzlibrary{shapes,arrows}
\usetikzlibrary{arrows.meta}
\usepackage{mathtools}
\usepackage{environ}
\makeatletter
\tikzset{%
  block/.style    = {draw, thick, rectangle, minimum height = 3em,
    minimum width = 3em},
  sum/.style      = {draw, circle, node distance = 2cm}, 
    input/.style    = {coordinate}, 
  output/.style   = {coordinate} 
}
\newsavebox{\measure@tikzpicture}
\NewEnviron{scaletikzpicturetowidth}[1]{%
  \def\tikz@width{#1}%
  \begin{lrbox}{\measure@tikzpicture}%
  \BODY
  \end{lrbox}%
  \pgfmathparse{#1/\wd\measure@tikzpicture}%
  \BODY
}
\usetikzlibrary{calc}
\def\centerarc[#1](#2)(#3:#4:#5)
    { \draw[#1] ($(#2)+({#5*cos(#3)},{#5*sin(#3)})$) arc (#3:#4:#5); }

\tikzset{
 thicker/.style={line width=#1\pgflinewidth},
 thicker/.default={2},
}
\def\mic(#1)(#2)(#3)(#4);{
    \begin{scope}[shift={(#1)},rotate={#2},scale={#3}]
        \draw[#4] (0,0) circle (0.22);
        \draw[thicker=1.5,#4] 
            (-0.22,-0.3) -- (-0.22,0.3);
    \end{scope}
}
\usetikzlibrary{shapes.geometric}
\def\speaker(#1)(#2)(#3);{
    \begin{scope}[shift={(#1)},rotate={#2},scale={#3}]
        \draw (-0.8,-0.40) rectangle (-0.35,0.40);
        \draw[-] (-0.35,0.40) -- (0,0.65) -- (0,-0.65) -- (-0.35,-0.40);
    \end{scope}
}
\usetikzlibrary{arrows}

\usetikzlibrary{shapes.multipart} 

\usepackage{empheq}

\definecolor{CBOrange}{HTML}{E69F00}
\definecolor{CBBlue}{HTML}{56B4E9}

\title{Incremental Averaging Method to Improve Graph-Based Time-Difference-of-Arrival Estimation}

\name{Klaus Br\"{u}mann$^{1}$,
      Kouei Yamaoka$^{2}$,
      Nobutaka Ono$^{3}$,
      Simon Doclo$^{1}$\thanks{This work was funded by the German Academic Exchange Service (DAAD), the Federal Ministry of Education and Research (BMBF) - Project ID 57711070, and the German Research Foundation (DFG) under Germany’s Excellence Strategy - EXC 2177/1 - Project ID 390895286. 
      }}
\address{$^{1}$Carl von Ossietzky Universit\"{a}t, Oldenburg, Germany \;
$^{2}$University of Tokyo, Tokyo, Japan \;\\
$^{3}$Tokyo Metropolitan University, Tokyo, Japan
}

\begin{document}

\maketitle

\begin{abstract}
Estimating the position of a speech source based on time-differences-of-arrival (TDOAs) is often adversely affected by background noise and reverberation. A popular method to estimate the TDOA between a microphone pair involves maximizing a generalized cross-correlation with phase transform (GCC-PHAT) function. Since the TDOAs across different microphone pairs satisfy consistency relations, generally only a small subset of microphone pairs are used for source position estimation. Although the set of microphone pairs is often determined based on a reference microphone, recently a more robust method has been proposed to determine the set of microphone pairs by computing the minimum spanning tree (MST) of a signal graph of GCC-PHAT function reliabilities. To reduce the influence of noise and reverberation on the TDOA estimation accuracy, in this paper we propose to compute the GCC-PHAT functions of the MST based on an average of multiple cross-power spectral densities (CPSDs) using an incremental method. In each step of the method, we increase the number of CPSDs over which we average by considering CPSDs computed indirectly via other microphones from previous steps. Using signals recorded in a noisy and reverberant laboratory with an array of spatially distributed microphones, the performance of the proposed method is evaluated in terms of TDOA estimation error and 2D source position estimation error. Experimental results for different source and microphone configurations and three reverberation conditions show that the proposed method considering multiple CPSDs improves the TDOA estimation and source position estimation accuracy compared to the reference microphone- and MST-based methods that rely on a single CPSD as well as steered-response power-based source position estimation.\end{abstract}\vspace*{-1 mm}

\section{Introduction} 
\vspace*{-0.5 mm}
\label{sec: Introduction} 
In various speech communication applications such as videoconferencing and smart speakers, microphone arrays are used to localize speech sources in noisy and reverberant environments, more in particular estimate their direction-of-arrival or position \cite{dibiase2001robust, huang2008time, madhu2008acoustic, brutti2010multiple, pertila2018multichannel}. 
Several model-based and learning-based methods have been proposed, 
e.g., methods using time-differences-of-arrival (TDOAs) \cite{chan1994simple, dibiase2001robust, gustafsson2003source, huang2001real, yang2020multiple, evers2020locata, brumann20223d}, the steered response power with phase transform (SRP-PHAT) method \cite{cobos2010modified, nunes2014steered, dietzen2021low, grinstein2024steered}, subspace-based methods \cite{schmidt1986multiple, dmochowski2007broadband}, 
and deep neural network-based methods \cite{grinstein2023neural, varzandeh2025improving}. 
In this paper, we focus on TDOA-based methods, where the source localization relies on prior estimation of the TDOAs. 
A few methods exist to estimate the TDOA between a given microphone pair \cite{benesty2000adaptive, dvorkind2005time}. 
Here we use the widely adopted method \cite{knapp1976generalized} based on maximizing the generalized cross-correlation with phase transforms (GCC-PHAT) function. 
However, noise and reverberation may introduce additional peaks in the GCC-PHAT function, affecting the accuracy of the estimated TDOAs and therefore also the source localization \cite{garcia2023exploiting}. 

Since consistency relations exist for TDOAs between multiple microphone pairs, generally only a small subset of microphone pairs, i.e., the minimal set \cite{scheuing2008correlation, velasco2016tdoa}, are used for source localization, to reduce the likelihood of including TDOA outliers. 
A simple, commonly used method to determine the minimal set of TDOAs is to relate them to a common reference microphone \cite{canclini2015robust, le2018rank}, where the choice of the reference microphone may have a large impact on the reliability of the GCC-PHAT functions \cite{gustafsson2003source}. 
Using graph theory, a more robust method has recently been proposed in \cite{yamaoka2023minimum} to determine the optimal minimal set in terms of GCC-PHAT reliability. 
This minimal set is determined as the minimum spanning tree (MST) of the signal graph and does not require choosing 
a common reference microphone. 
Nevertheless, even though the MST method has been shown to be more robust in terms of avoiding TDOA outliers, strong noise or reverberation in one or more microphones may mean that it is not possible to avoid TDOA outliers in the minimal set of microphone pairs. 

Instead of computing each GCC-PHAT function for each microphone pair in the MST based on a single cross-power spectral density (CPSD), in this paper, we propose an incremental method to average over multiple CPSDs from multiple microphone pairs. 
Addressing a different microphone pair of the MST-based minimal set in each step, we order the microphones based on their GCC-PHAT reliabilities and incrementally re-estimate the TDOAs using the averaged CPSDs. 
In each step, we propose to incorporate an additional CPSD, estimated indirectly via another microphone which was used in previous steps, which generally has a similar, or more reliable GCC-PHAT function. 
To include these additional CPSDs into the average requires phase alignment, which is accomplished using a phase shift based on TDOAs which were re-estimated in previous steps. 

The performance of the proposed TDOA estimation method is compared to baseline reference microphone- and MST-based methods, as well as the steered-response power-based position estimation, in terms of TDOA estimation error and source 2D position estimation error. 
Based on recorded noisy and reverberant speech signals, experimental results for three reverberation conditions demonstrate that the proposed TDOA estimation method outperforms all of the considered baseline methods in a range of reverberant environments, demonstrating that it is beneficial to base TDOA estimation on an average of multiple CPSDs instead of just a single CPSD, which is possible through our incremental method. \vspace*{-3.5 mm}
\section{State-of-the-Art TDOA Estimation} 
\vspace*{-1.5 mm}
\label{sec: State-of-the-Art} 
We consider a noisy and reverberant acoustic environment with a single speech source at position $\pbf$ and a microphone array with $M$ microphones at positions $\left[ \m_{1}, \dots, \m_{M} \right] \in \Real^{P\times M}$, where $P$ is the dimensionality of the acoustic scenario. 
Assuming synchronized microphones and free-field transmission, i.e., no objects between the source and the microphones, the TDOA between microphones $i$ and $j$ is given by $\tau_{i,j}(\pbf) = (||\pbf-\m_{i}||_2 - ||\pbf-\m_{j}||_2) / \nu$, 
where $\nu$ denotes the speed of sound. 
The $M \times M$-dimensional TDOA matrix, containing the TDOAs between all microphone pairs, is defined as $\T(\pbf) = [\tau_{i,j}(\pbf)]$, and is an anti-symmetric matrix with rank 2 \cite{velasco2016tdoa}. 
Since the TDOAs across different microphone pairs satisfy consistency relations (i.e., $\tau_{i,j} = \tau_{i,m} - \tau_{j,m} \; , \; \forall m$), it has been shown in \cite{so2008closed,velasco2016tdoa} that TDOA matrices can be written as 
\begin{equation}
\T(\pbf) \; = \; \taubf_{m}^{}(\pbf) \1_{M}^{\textrm{T}} - \1_{M} \taubf_{m}^{\textrm{T}}(\pbf) \quad \forall \;\; m \; ,
\label{eq: consistency}
\end{equation}
with $\taubf_{m}(\pbf) = [\tau_{1,m}(\pbf), \dots, \tau_{M,m}(\pbf)]^{\textrm{T}}$ the $M$-dimensional vector of TDOAs relative to the $m$-th reference microphone (with $\tau_{m,m}(\pbf) = 0$) and $\1_{M}$ an $M$-dimensional vector of ones. 

A commonly used method to estimate the TDOA $\tau_{i,j}(\pbf)$ is based on the GCC-PHAT function \cite{knapp1976generalized} between microphones $i$ and $j$, defined as\vspace*{-1 mm} 
\begin{equation}
    \xi_{i,j}(\tau) \; = \int_{-\infty}^{\infty} {\phi}_{i,j}^{}(\omega) \; \exp{(\ji \omega \tau)} d\omega \; ,
    \label{eq: Time-Domain GCC-PHAT}
\end{equation}
with imaginary number $\ji = \sqrt{-1}$, radial frequency $\omega$ and time lag $\tau$. 
The phase transform (PHAT)-weighted CPSD ${\phi}_{i,j}(\omega)$ is given by 
\begin{equation}
    {\phi}_{i,j}^{}(\omega) 
    \; = \; 
    \frac{{\psi}_{i,j}(\omega)}{| {\psi}_{i,j}(\omega)|}
    \; ,
\label{eq: psi}
\end{equation}
where ${\psi}_{i,j}(\omega) = \EX\{ Y_{i}^{}(\omega) Y_{j}^{*}(\omega) \}$ denotes the CPSD between the microphones $i$ and $j$, $Y_{m}(\omega)$ denotes the $m$-th microphone signal in the continuous-time Fourier transform domain, and $\EX\{\cdot{}\}$ denotes the expectation operator. 
The estimated TDOA $\tilde{\tau}_{i,j}$ is computed as the time lag that maximizes $\xi_{i,j}(\tau)$, i.e., 
\begin{equation}
    \tilde{\tau}_{i,j} \;\; = \;\; \argmax_{\tau} \;\; \xi_{i,j}(\tau) \; . 
    \label{eq: TDOA estimation}
\end{equation}
Due to noise and reverberation, the GCC-PHAT functions may exhibit additional peaks, which result in TDOA estimation errors if they are higher than the peak corresponding to the direct path signal \cite{garcia2023exploiting}. 
Due to estimation errors, the resulting estimated TDOA matrix is not guaranteed to be consistent due to estimation errors. 
To avoid discrepancies between inconsistent estimated TDOAs, various methods have proposed to use only the estimated TDOAs between a minimal set of $M-1$ microphone pairs, from which a consistent estimated TDOA matrix can be constructed. 
In the following subsections, we will discuss two classes of methods to determine the minimal set of microphone pairs, either based on graph theory in \cref{sec: MST} or using a common reference microphone in \cref{sec: Reference Methods}. \vspace*{-1 mm}

\subsection{Minimal Set Based on MST}
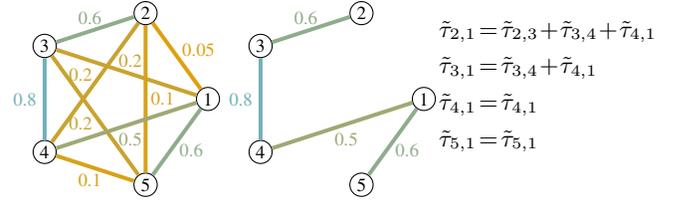
\begin{figure}[t]
    \centering
\centerline{
\begin{minipage}{0.24\linewidth} 
\scriptsize{}
\begin{tikzpicture}[scale=1.2] 

    \foreach \i in {1,...,5} {
        \node[circle, draw, fill=white, inner sep=1pt, minimum size=3mm] (\i) at ({72*(\i-1)}:1) {\i};
    }

    \foreach \i in {1,...,5} {
        \foreach \j in {\i,...,5} {
            \ifnum\i<\j
                \draw[color=white] (\i) -- (\j);
            \fi
        }
    }

    \draw[color=CBBlue!10!CBOrange,line width=1.5 pt] (4) -- (5) node[midway, below, font=\scriptsize, xshift=-2pt, yshift=1pt,color=CBBlue!10!CBOrange] {0.1}; 
    \draw[color=CBBlue!05!CBOrange,line width=1.5 pt] (2) -- (1) node[midway, above right, font=\scriptsize, xshift=-1pt, yshift=-3pt,color=CBBlue!05!CBOrange] {0.05}; 
    \draw[color=CBBlue!20!CBOrange,line width=1.5 pt] (3) -- (1) node[midway, above, font=\scriptsize, xshift=1.5pt, yshift=-1pt,color=CBBlue!20!CBOrange] {0.2}; 
    \draw[color=CBBlue!20!CBOrange,line width=1.5 pt] (2) -- (4) node[midway, left, font=\scriptsize, xshift=2pt, yshift=3pt,color=CBBlue!20!CBOrange] {0.2}; 
    \draw[color=CBBlue!10!CBOrange,line width=1.5 pt] (2) -- (5) node[midway, right, font=\scriptsize, xshift=-1pt, yshift=0pt,color=CBBlue!10!CBOrange] {0.1}; 
    \draw[color=CBBlue!20!CBOrange,line width=1.5 pt] (3) -- (5) node[midway, left, font=\scriptsize, xshift=2pt, yshift=-3pt,color=CBBlue!20!CBOrange] {0.2}; 

    \draw[color=CBBlue!80!CBOrange,line width=1.5 pt] (3) -- (4) node[midway, left, font=\scriptsize, xshift=0pt, yshift=0pt] {0.8}; 
    \draw[color=CBBlue!50!CBOrange,line width=1.5 pt] (4) -- (1) node[midway, below, font=\scriptsize, xshift=1.5pt, yshift=0.5pt] {0.5}; 
    \draw[color=CBBlue!60!CBOrange,line width=1.5 pt] (2) -- (3) node[midway, above, font=\scriptsize, xshift=-2pt, yshift=-1pt] {0.6}; 
    \draw[color=CBBlue!60!CBOrange,line width=1.5 pt] (1) -- (5) node[midway, right, font=\scriptsize, xshift=-2pt, yshift=-3pt] {0.6}; 

\end{tikzpicture}
\end{minipage}
\hfill 
\begin{minipage}{0.24\linewidth} 
\scriptsize{}
\begin{tikzpicture}[scale=1.2]
    \foreach \i in {1,...,5} {
        \node[circle, draw, fill=white, inner sep=1pt, minimum size=3mm] (\i) at ({72*(\i-1)}:1) {\i};
    }
    \draw[color=CBBlue!80!CBOrange,line width=1.5 pt] (3) -- (4) node[midway, left, font=\scriptsize, xshift=0pt, yshift=0pt] {0.8}; 
    \draw[color=CBBlue!50!CBOrange,line width=1.5 pt] (4) -- (1) node[midway, below, font=\scriptsize, xshift=1.5pt, yshift=0.5pt] {0.5}; 
    \draw[color=CBBlue!60!CBOrange,line width=1.5 pt] (2) -- (3) node[midway, above, font=\scriptsize, xshift=-2pt, yshift=-1pt] {0.6}; 
    \draw[color=CBBlue!60!CBOrange,line width=1.5 pt] (1) -- (5) node[midway, right, font=\scriptsize, xshift=-2pt, yshift=-3pt] {0.6}; 
    
    
\end{tikzpicture}
\end{minipage}
\hfill 
\begin{minipage}{0.24\linewidth} 
\centering
\begin{align*}
\tilde{\tau}_{2,1} &= \tilde{\tau}_{2,3} + \tilde{\tau}_{3,4} + \tilde{\tau}_{4,1}\\
\tilde{\tau}_{3,1} &= \tilde{\tau}_{3,4} + \tilde{\tau}_{4,1}\\
\tilde{\tau}_{4,1} &= \tilde{\tau}_{4,1}\\
\tilde{\tau}_{5,1} &= \tilde{\tau}_{5,1}
\end{align*}\vspace*{2 mm}
\end{minipage}
}
    \caption{Left: Exemplary graph of GCC-PHAT reliability values between $M=5$ microphones. Middle: Corresponding minimum spanning tree. Right: Corresponding TDOAs, rewritten relative to the first reference microphone.}
    \label{fig: MST}
\end{figure}

\vspace*{-0.5 mm}
\label{sec: MST}
Using graph theory, a method was recently proposed in \cite{yamaoka2023minimum} to determine the minimal set of microphone pairs based on GCC-PHAT reliability. 
The undirected signal graph $G = (\mathcal{M},{\Theta})$ is characterized by the set of vertices $\mathcal{M}=\{1,\dots,M\}$, representing the microphones, and the set of edges $\Theta = \{ \, i \, , \, j \; | \; i \, , \, j \; \in \mathcal{M} \; \cap \; i \, \neq \, j \,\}$ (corresponding to the cost, defined as the negative reliability). 
Similarly as in \cite{yamaoka2023minimum}, reliability is defined in this paper as the maximum value of the GCC-PHAT function \cref{eq: Time-Domain GCC-PHAT}, i.e.,\vspace*{-2 mm} 
\begin{equation}
R_{i,j} \; = \; \xi_{i,j}(\tilde{\tau}_{i,j}) \; . 
    \label{eq: Reliability}\vspace*{-1 mm}
\end{equation}
In the MST method, the minimal set of microphone pairs is determined by the edges ${\Theta}^{\textrm{MST}} \subseteq \Theta$, which are chosen by the Prim method \cite{dijkstra1959note} such that $G^{\textrm{MST}} = (\mathcal{M}, {\Theta}^{\textrm{MST}})$ forms a spanning tree and the total negative reliability, i.e., $\sum_{i,j\subseteq {\Theta}^{\textrm{MST}}} -R_{i,j}$ is minimized.
As described in \cite{yamaoka2023minimum}, the set of TDOAs corresponding to these microphone pairs is then rewritten relative to an arbitrarily chosen reference microphone in a TDOA vector $\tilde{\taubf}_{m}$, which is generally how they are used for TDOA-based position estimation. 
An exemplary MST and the corresponding set of TDOAs are shown in \cref{fig: MST}. 

\vspace*{-1 mm}

\subsection{Minimal Set Based on Common Reference Microphone} 
\vspace*{-0.5 mm}
\label{sec: Reference Methods} 
A commonly used minimal set of microphone pairs for TDOA estimation is the set of microphone pairs relative to a selected reference microphone. 
In this paper, we consider three baseline methods for choosing the reference microphone index $m$. 
In the arbitrary method (Ref-A), the reference microphone is chosen randomly, i.e., $\hat{m}_{\textrm{A}} \sim \mathcal{U}(\{1, 2, \dots, M\})$. 
In the centroid method (Ref-C), the reference microphone is chosen as the microphone that is closest to the centroid of the microphone array as in \cite{dibiase2001robust}, i.e., $\hat{m}_{\textrm{C}} \; = \; \argmin_{m} \; {||\m_m - \frac{1}{M}\sum_{i=1}^{M} \m_{i} ||_{2}}$. 
In the reliability-based method (Ref-R), the reference microphone is chosen based on the GCC-PHAT reliability. 
Rather than selecting the reference microphone based on the average GCC-PHAT reliability between this microphone and all other microphones, we propose to select the reference microphone for which the minimum GCC-PHAT reliability between this microphone and all other microphones is highest, i.e., 
\begin{equation}
\hat{m}_{\textrm{R}} \; = \; \argmax_{m}\{ \argmin_{i \neq m}{ \;R_{i,m}} \; \} \; .
\label{eq: Ref-R}\vspace*{-1 mm}
\end{equation}
This reduces the risk that a highly erroneous estimated TDOA is included in the minimal set. \vspace*{-2.5 mm}

\section{TDOA Estimation by Incrementally Averaging Multiple CPSDs}\vspace*{-1.5 mm}
\label{sec: RMST}
Although the MST-based minimal set of microphone pairs reduces the chances of including a TDOA outlier compared to other minimal sets, it should be realized that each TDOAs is computed from a single PHAT-weighted CPSD. 
Therefore, including microphone pairs with TDOA outliers may be unavoidable when one or more microphones are subject to high levels of noise and reverberation. 
To further improve the accuracy of the estimated TDOAs, in this section we propose an incremental method to re-estimate the TDOAs corresponding to the MST based on an average of multiple CPSDs from multiple microphone pairs. 
This averaging of CPSDs is based on a directly computed CPSD and indirectly estimated CPSDs which are computed via other microphones from previous steps. 
After introducing indirect CPSD estimation in \cref{sec: Indirect CPSD Estimation}, in \cref{sec: RMST Method} we describe the proposed incremental method for averaging CPSDs.\vspace*{-1 mm}

\subsection{Indirect CPSD Computation}\label{sec: Indirect CPSD Estimation}\vspace*{-0.5 mm}
Considering only the direct source component (i.e., in noise-free and anechoic conditions) the CPSD between microphones $i$ and $j$ can be written as ${\psi}_{i,j}(\omega) = \exp(-\ji \omega \tau_{i,j}(\pbf))/(16\pi^{2}d_{i}d_{j})$, with $d_{i} = ||\pbf - \m_{i}||_{2}$ the distance between the source and the $i$-th microphone \cite{madhu2008acoustic, huang2008time, pertila2018multichannel}. 
Since for any microphone $k$ it can be easily shown that $\exp(-\ji \omega \tau_{i,j}(\pbf)) \; = \; \exp(-\ji \omega \tau_{i,k}(\pbf)) \; \exp(\ji \omega \tau_{j,k}(\pbf))$, the CPSD between microphones $i$ and $j$ can be indirectly computed by applying a phase shift to the CPSD between microphones $i$ and $k$ and compensating for the distance-related attenuation, similarly to \cite{brumann2024steered}, i.e., \vspace*{-3 mm}
\begin{equation}
{\psi}_{i,j}^{}(\omega) \; = \; 
\frac{d_{k}}{d_{j}} \; {\psi}_{i,k}^{}(\omega) \; \exp(\ji \omega {\tau}_{j,k}(\pbf)) \; .
\label{eq: psi RMST theory}
\end{equation}
In practice, the distances $d_{j}$ and $d_{k}$ are of course unavailable and are difficult to estimate. 
Since we are mainly interested in the phase, we assume $d_{j} = d_{k}$, and approximate the indirectly computed CPSD as 
\begin{equation}
\boxed{
\tilde{\psi}_{i,j}^{[k]}(\omega) \; = \; {\psi}_{i,k}^{}(\omega) \; \exp(\ji \omega {\tau}_{j,k}(\pbf)) 
}
\label{eq: psi RMST}
\end{equation}
\vspace*{-2 mm}

\subsection{Incremental Method for Averaging CPSDs}\label{sec: RMST Method}\vspace*{-1 mm}
Based on the vertices and edges of the MST, the proposed method builds up a consistent TDOA matrix with re-estimated TDOAs from scratch in $M-1$ steps, where $H$ denotes the step index. 
To improve the estimation of CPSDs corresponding to less reliable GCC-PHAT functions in later steps, we begin with CPSDs corresponding to the highest GCC-PHAT reliabilities. 
Therefore, before introducing the method, the edges are ordered based on reliability 
for which we define an $M-1$-dimensional vector $\rbf$ of reliabilities and a corresponding $(M-1) \times 2$-dimensional matrix $\V$ containing $M-1$ rows of MST edges $\left[i,j\right] , \; i,j \in {\Theta}^{\textrm{MST}}$. 
The first entry of $\rbf$ contains the highest reliability value of the MST and the first row of $\V$ contains the pair of vertices connected by the associated edge. 
Subsequent entries of $\rbf$ can only correspond to an edge connected to one of the vertices from previous rows of $\V$, in order of descending reliability. 
For the example in \cref{fig: MST}, the reliability vector would be $\rbf_{\textrm{MST}} = [0.8, 0.6, 0.5, 0.6]^{\textrm{T}}$ and the rows of $\V$ would be $[4,3]$, $[3,2]$, $[4,1]$, and $ [5,1]$. 
The indices $i[H], \;\; j[H]$ refer to the microphones in the $H$-th row of $\V$. 
In each step $H$, the proposed method (referred to as MST+) re-estimates the TDOA between the microphones $i[H]$ and $j[H]$. 

In the first step ($H=1$), the TDOA between $i(1)$ and $j(1)$ is estimated based on the same GCC-PHAT function \cref{eq: Time-Domain GCC-PHAT} as used for computing the reliability in \cref{eq: Reliability}. 
In other words, no TDOA re-estimation is performed. 
In subsequent steps ($H>1$), we propose to re-estimate the TDOAs by incorporating knowledge from previous steps. 
Considering noisy and reverberant speech signals, we assume that the noise and reverberation destructively interfere across multiple microphone pairs, in contrast to the direct path component. 
Therefore, we propose to average out the spurious GCC-PHAT function peaks, corresponding to TDOA outliers, by averaging over the directly computed CPSD ${{\psi}}_{i[H],j[H]}^{}(\omega)$ and $H-1$ indirectly computed CPSDs via all microphones used in previous steps. 
As such, the PHAT-weighted CPSD between microphones $i[H]$ and $j[H]$ is computed as\vspace*{-1.5 mm}
\begin{equation}
\boxed{
    \tilde{{\phi}}_{i[H],j[H]}^{'}(\omega) \; = \; \frac{ {{\psi}}_{i[H],j[H]}^{}(\omega) \; + \; \sum_{h=1}^{H-1} \; \tilde{{\psi}}_{i[H],j[H]}^{[j[h]]}(\omega)
    }{
    | {{\psi}}_{i[H],j[H]}^{}(\omega) \; + \; \sum_{h=1}^{H-1} \; \tilde{{\psi}}_{i[H],j[H]}^{[j[h]]}(\omega) |} 
    }
    \label{eq: Reinforced CPSD}
\end{equation}
where, based on \cref{eq: psi RMST}, the CPSDs in \cref{eq: Reinforced CPSD} are computed using TDOAs $\tilde{\tau}_{j[H],j[h]}'$ which were re-estimated in previous steps $j[h]$ as 
\begin{equation}
\tilde{{\psi}}_{i[H],j[H]}^{[j[h]]}(\omega) \; = \; 
{\psi}_{i[H],j[h]}^{}(\omega)  \; \exp(\ji \omega \tilde{\tau}_{j[H],j[h]}')  \;\; , \quad  \; 1 \leq h < H \;.
\label{eq: psi RMST est}
\end{equation}
By incorporating CPSDs indirectly estimated via microphones corresponding to edges with generally higher GCC-PHAT reliabilities, it is expected that, in addition to mitigating spurious peaks corresponding to TDOA outliers, also the TDOA estimation accuracy can be improved. 
The TDOA $\tilde{\tau}_{i[H],j[H]}^{'}$ is then re-estimated as the time lag maximizing the GCC-PHAT function in \cref{eq: Time-Domain GCC-PHAT} using the PHAT-weighted CPSD in \cref{eq: Reinforced CPSD}. 
To use the re-estimated TDOAs for the phase alignment of the indirectly estimated CPSDs in \cref{eq: psi RMST est} in subsequent steps, we re-estimate additional TDOAs for each of the microphones used in previous steps $j[h]$ as \vspace*{-2 mm}
\begin{equation}
\tilde{\tau}_{i[h],j[H]}^{'} \; = \; \tilde{\tau}_{i[h],j[h]}^{'} - \tilde{\tau}_{j[H],j[h]}^{'} \;\; , \quad  \; 1 \leq h < H \;.
\label{eq: RMST TDOA complete}
\end{equation}
These steps are repeated until all TDOAs corresponding to the MST have been re-estimated. 
For the example in \cref{fig: MST}, \cref{fig: RMST} shows graphs and the relevant entries of the estimated TDOA matrix. 
\usetikzlibrary{patterns,patterns.meta}
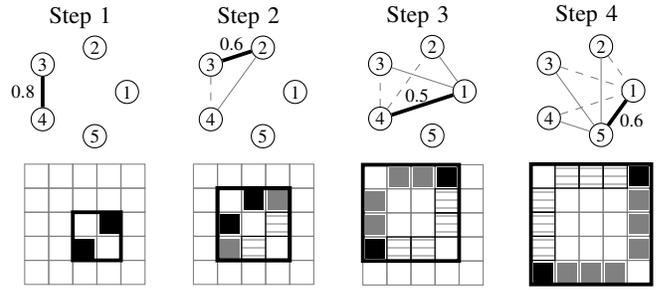
\begin{figure}
    \centering
\begin{minipage}{0.24\linewidth}
\centering
\begin{tikzpicture}[scale=0.62]
\draw (0,1.6) node {Step 1};
\scriptsize{}\foreach \i in {1,...,5} {
        \node[circle, draw, fill=white, inner sep=1pt, minimum size=3mm] (\i) at ({72*(\i-1)}:1) {\i};
    }
    \draw[draw=none,line width=0.5 mm] (2) -- (3) node[midway, above, font=\scriptsize, xshift=-2pt, yshift=-1pt] {\hphantom{0.5}};  
    \draw[line width=0.5 mm] (3) -- (4) node[midway, left, font=\scriptsize, xshift=0pt, yshift=0pt] {0.8};
    \draw[draw=none,line width=0.5 mm] (1) -- (1) node[midway, right, font=\scriptsize, xshift=0pt, yshift=0pt] {\hphantom{0.8}}; 
\end{tikzpicture}\vspace*{2 mm}
\begin{tikzpicture}[scale=0.32]
\foreach \x in {1, 2, 3, 4, 5} {
    \foreach \y in {1, 2, 3, 4, 5} {
        \draw[gray] (\y-1,-\x+1) rectangle (\y,-\x);
    }
}
\fill[black] (2.1,-3.1) rectangle (2.9,-3.9);
\fill[black] (3.1,-2.1) rectangle (3.9,-2.9);
\draw[line width = 0.50 mm] (2,-2) rectangle (4,-4);

\end{tikzpicture}
\end{minipage}\hfill
\begin{minipage}{0.24\linewidth}
\centering
\begin{tikzpicture}[scale=0.62]
\draw (0,1.6) node {Step 2};
\scriptsize{}\foreach \i in {1,...,5} {
        \node[circle, draw, fill=white, inner sep=1pt, minimum size=3mm] (\i) at ({72*(\i-1)}:1) {\i};
    }
    \draw[line width=0.5 mm] (2) -- (3) node[midway, above, font=\scriptsize, xshift=-2pt, yshift=-1pt] {{0.6}};
    \draw[draw=none,line width=0.5 mm] (1) -- (1) node[midway, right, font=\scriptsize, xshift=0pt, yshift=0pt] {\hphantom{0.8}}; 
    \draw[gray] (4) -- (2);
    \draw[gray,dashed] (3) -- (4) node[midway, left, font=\scriptsize, xshift=0pt, yshift=0pt] {\hphantom{0.8}};
\end{tikzpicture}\vspace*{2 mm}
\begin{tikzpicture}[scale=0.32]
\scriptsize{}\foreach \x in {1, 2, 3, 4, 5} {
    \foreach \y in {1, 2, 3, 4, 5} {
        \draw[gray] (\y-1,-\x+1) rectangle (\y,-\x);
    }
}
\draw[pattern={Lines[angle=45]},pattern color=gray]
 (2,-3) rectangle +(1,-1);
 \draw[pattern={Lines[angle=45]},pattern color=gray]
 (3,-2) rectangle +(1,-1);

\fill[gray] (1.1,-3.1) rectangle (1.9,-3.9);
\fill[gray] (3.1,-1.1) rectangle (3.9,-1.9);
\fill[black] (1.1,-2.1) rectangle (1.9,-2.9);
\fill[black] (2.1,-1.1) rectangle (2.9,-1.9);
\draw[line width = 0.55 mm] (1,-1) rectangle (4,-4);

\end{tikzpicture}
\end{minipage}\hfill
\begin{minipage}{0.24\linewidth}
\centering
\begin{tikzpicture}[scale=0.62]
\draw (0,1.6) node {Step 3};
\scriptsize{}\foreach \i in {1,...,5} {
        \node[circle, draw, fill=white, inner sep=1pt, minimum size=3mm] (\i) at ({72*(\i-1)}:1) {\i};
    }
    \draw[line width=0.5 mm] (4) -- (1) node[midway, above, font=\scriptsize, xshift=-2pt, yshift=-2pt] {0.5}; 
    \draw[draw=none,line width=0.5 mm] (3) -- (4) node[midway, left, font=\scriptsize, xshift=0pt, yshift=0pt] {\hphantom{0.5}}; 
    \draw[draw=none,line width=0.5 mm] (1) -- (1) node[midway, right, font=\scriptsize, xshift=0pt, yshift=0pt] {\hphantom{0.5}}; 
    \draw[gray] (3) -- (1) ; 
    \draw[gray] (2) -- (1) ; 
    \draw[gray,dashed] (4) -- (3); 
    \draw[gray,dashed] (4) -- (2); 
\end{tikzpicture}\vspace*{2 mm}
\begin{tikzpicture}[scale=0.32]
\foreach \x in {1, 2, 3, 4, 5} {
    \foreach \y in {1, 2, 3, 4, 5} {
        \draw[gray] (\y-1,-\x+1) rectangle (\y,-\x);
    }
}


\fill[black] (3.1,-0.1) rectangle (3.9,-0.9);
\fill[black] (0.1,-3.1) rectangle (0.9,-3.9);
\fill[gray] (0.1,-1.1) rectangle (0.9,-1.9);
\fill[gray] (1.1,-0.1) rectangle (1.9,-0.9);
\fill[gray] (0.1,-2.1) rectangle (0.9,-2.9);
\fill[gray] (2.1,-0.1) rectangle (2.9,-0.9);
\draw[pattern={Lines[angle=45]},pattern color=gray]
 (2,-3) rectangle +(1,-1);
\draw[pattern={Lines[angle=45]},pattern color=gray]
 (3,-2) rectangle +(1,-1);
\draw[pattern={Lines[angle=45]},pattern color=gray]
 (2-1,-4+1) rectangle +(1,-1);
\draw[pattern={Lines[angle=45]},pattern color=gray]
 (4-1,-2+1) rectangle +(1,-1);
\draw[line width = 0.50 mm] (0,0) rectangle (4,-4);

\end{tikzpicture}
\end{minipage}\hfill
\begin{minipage}{0.24\linewidth}
\centering
\begin{tikzpicture}[scale=0.62]
\draw (0,1.6) node {Step 4};
\scriptsize{}\foreach \i in {1,...,5} {
        \node[circle, draw, fill=white, inner sep=1pt, minimum size=3mm] (\i) at ({72*(\i-1)}:1) {\i};
    }
    \draw[line width=0.5 mm] (1) -- (5) node[midway, right, font=\scriptsize, xshift=-2pt, yshift=-3pt] {0.6}; 
    \draw[draw=none,line width=0.5 mm] (3) -- (4) node[midway, left, font=\scriptsize, xshift=0pt, yshift=0pt] {\hphantom{0.5}};  
    \draw[draw=none,line width=0.5 mm] (1) -- (1) node[midway, right, font=\scriptsize, xshift=0pt, yshift=0pt] {\hphantom{0.5}}; 
    \draw[] (5) -- (1) ; 
    \draw[gray] (5) -- (2) ; 
    \draw[gray] (5) -- (3) ; 
    \draw[gray] (5) -- (4) ; 
    \draw[gray,dashed] (3) -- (1); 
    \draw[gray,dashed] (2) -- (1); 
    \draw[gray,dashed] (4) -- (1); 
\end{tikzpicture}\vspace*{2 mm}
\begin{tikzpicture}[scale=0.32]
\foreach \x in {1, 2, 3, 4, 5} {
    \foreach \y in {1, 2, 3, 4, 5} {
        \draw[gray] (\y-1,-\x+1) rectangle (\y,-\x);
    }
}



\fill[black] (4.1,-0.1) rectangle (4.9,-0.9);
\fill[black] (0.1,-4.1) rectangle (0.9,-4.9);
\fill[gray] (4.1,-1.1) rectangle (4.9,-1.9);
\fill[gray] (1.1,-4.1) rectangle (1.9,-4.9);
\fill[gray] (4.1,-2.1) rectangle (4.9,-2.9);
\fill[gray] (2.1,-4.1) rectangle (2.9,-4.9);
\fill[gray] (4.1,-3.1) rectangle (4.9,-3.9);
\fill[gray] (3.1,-4.1) rectangle (3.9,-4.9);
 \draw[pattern={Lines[angle=45]},pattern color=gray]
 (4-1,-1+1) rectangle +(1,-1);
\draw[pattern={Lines[angle=45]},pattern color=gray]
 (1-1,-4+1) rectangle +(1,-1);
\draw[pattern={Lines[angle=45]},pattern color=gray]
 (1-1,-2+1) rectangle +(1,-1);
\draw[pattern={Lines[angle=45]},pattern color=gray]
 (2-1,-1+1) rectangle +(1,-1);
 \draw[pattern={Lines[angle=45]},pattern color=gray]
 (1-1,-3+1) rectangle +(1,-1);
\draw[pattern={Lines[angle=45]},pattern color=gray]
 (3-1,-1+1) rectangle +(1,-1);
\draw[line width = 0.50 mm] (0,0) rectangle (5,-5);

\end{tikzpicture}
\end{minipage}
    \caption{Graphs and TDOA matrices for all steps of the proposed incremental TDOA re-estimation method, corresponding to the example in \cref{fig: MST}. 
    The black lines and boxes correspond to the microphone pair $i[H], \;\; j[H]$ used in step $H$, the solid gray lines and boxes correspond to the microphone pairs used for indirect CPSD estimation and the dashed gray lines and boxes correspond to microphone pairs used for phase alignment using TDOAs re-estimated in previous steps (i.e., entries within the black border from a previous step).%
    }\label{fig: RMST}\end{figure}
\vspace*{-1 mm}
\section{Experimental Evaluation}\vspace*{-0.5 mm}
\label{sec: Evaluation}
Using real-world noisy and reverberant speech signals recorded using an array of spatially distributed microphones, in this section we compare the performance of the proposed MST+ method with the baseline MST and reference microphone-based TDOA estimation methods and the steered-response power method, both in terms of TDOA estimation error as well as source position estimation error. 
\Cref{sec: Scenario} outlines the acoustic scenario, in \cref{sec: Practical Implementation}, we discuss practical considerations for the implementation of the considered methods, and in \cref{sec: Performance Comparison} we discuss the experimental results.\vspace*{-0.5 mm} 

\subsection{Acoustic Scenario}\label{sec: Scenario}
\begin{figure}
    \centering
    \includegraphics[width=0.7\linewidth]{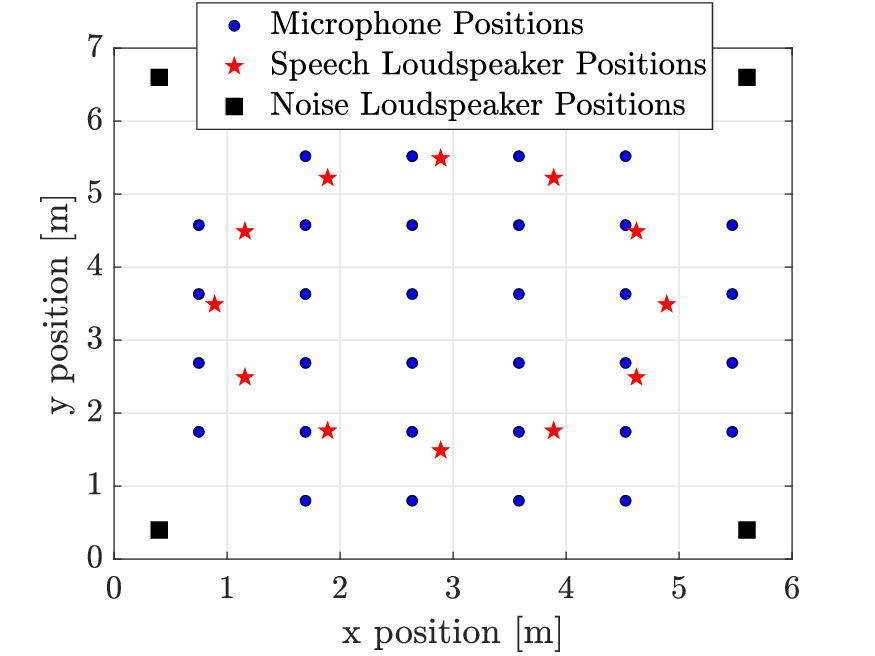}
    \caption{Layout of the distributed microphones ({\textcolor{blue}{\Huge\raisebox{-0.47ex}\textbullet}}\color{black}) and speech loudspeakers ({\color{red}{\normalsize$\bigstar$}}\color{black}) in the BRUDEX laboratory}
    \label{fig: BRUDEX}
\end{figure}

\vspace*{-0.5 mm}
In our experiments, we used noisy and reverberant signals from the BRUDEX database \cite{fejgin2023brudex}, where the speech source and the noise were recorded separately by $M=6$ distributed microphones in a laboratory with dimensions of about 6 m $\times$ 7 m $\times$ 2.7 m, whose layout can be seen in \cref{fig: BRUDEX}. 
A 10 s recorded speech signal randomly chosen from one of the three available recorded signals was considered for 120 different static source-microphone configurations (comprising 10 random microphone array configurations for each of the 12 available source positions) at a height of approximately 1.5 m. 
We considered three reverberation levels, corresponding to low ($T_{60} \approx 310$ ms), medium ($T_{60} \approx 510$ ms), and high ($T_{60} \approx 1300$ ms) reverberation. 
Diffuse-like babble noise produced by loudspeakers in the corners of the laboratory was added to the recorded speech signals at a signal-to-noise ratio of 5 dB averaged across the microphones. 
  
\vspace*{-0.5 mm}

\subsection{Practical Implementation}\label{sec: Practical Implementation}\vspace*{-0.5 mm}
The algorithms were implemented with a sampling frequency $f_s = 16$ kHz and a short-time Fourier transform (STFT) framework was used with a frame length of $1024$ samples (corresponding to 64 ms), 50\% overlap between frames, and a square-root-Hann analysis window.

The CPSDs between the microphone signals were calculated in the STFT domain using recursive smoothing as\vspace*{-1 mm} 
\begin{equation}
    {\psi}_{i,j}[k,l] \;\; = \;\; \lambda{\psi}_{i,j}[k,l-1] + (1-\lambda)Y_{i}^{}[k,l]Y_{j}^{*}[k,l] \; ,
    \label{eq: phase spectrum adaptive}
\end{equation}
where $k$ denotes the frequency bin index, and $l$ denotes the time frame index. 
We used a recursive smoothing factor $\lambda = 0.98$, corresponding to 1.58 s. 
The PHAT-weighted CPSD was computed as ${\phi}_{i,j}[k,l] = {\psi}_{i,j}[k,l]/|{\psi}_{i,j}[k,l]|$. 
The GCC-PHAT functions (based on an average of multiple CPSDs) were then computed using the inverse discrete Fourier transform for discrete time lags $n=\tau f_s$, i.e.,\vspace*{-1 mm} 
\begin{equation}
    \xi_{i,j}[n,l] \; = \; \frac{1}{(K-1)} \sum_{k=1}^{K-1} {\phi}_{i,j}^{}[k,l] \; \exp{\left(\frac{\ji 2\pi n k }{ K} \right)} \;\; ,
    \label{eq: IDFT of GCC-PHAT}
\end{equation} 
where $K$ denotes the DFT length. 
To achieve a more precise TDOA estimate, the GCC-PHAT functions 
were interpolated with an upsampling factor $R = 10$. 
We only considered physically realistic time-lags, i.e., $n_{i,j}^{\textrm{min}} \leq n_{i,j}^{}\leq n_{i,j}^{\textrm{max}}$, with $n_{i,j}^{\textrm{max}} = -n_{i,j}^{\textrm{min}} = R f_s D_{i,j} / \nu $, with $D_{i,j} = ||\m_{i}-\m_{j}||_{2}$ denoting the distance between the microphones $i$ and $j$. 
The TDOA estimate for the $l$-th time frame was then obtained as $\hat{\tau}_{i,j}[l] = \hat{n}_{i,j}[l]/(R f_s)$. 

To evaluate the usability of the estimated TDOAs from different methods for source position estimation, we used the spherical (SI) interpolation-based source position estimation method \cite{dibiase2001robust}. 
In each time-frame, the SI cost function (constrained to a distance of $5$ m from the centre of the room
) was iteratively solved with gradient descent. 
For comparison, we also considered the SRP-PHAT method used in \cite{brumann20223d}, i.e., the SRP-PHAT functional was evaluated on a 2D grid, first at a resolution of 10 cm, then at a resolution of 1 cm in the vicinities of the three grid points with the highest SRP-PHAT functional values, to find the global functional maximum. 
\vspace*{-0.5 mm} 

\subsection{Comparison of TDOA and Position Estimation Performance}\label{sec: Performance Comparison}\vspace*{-0.5 mm} 
To evaluate the TDOA estimation performance of the considered methods, we used the mean TDOA estimation error\vspace*{-1 mm} 
\begin{equation}
    \overline{\sigma} \;\; = \;\; \frac{1}{S(M-1)} \sum_{s=1}^{S} \sum_{m=2}^{M} \; |\hat{\tau}_{m,1}[s] - {\tau}_{m,1}(\pbf[s])| \; ,
\end{equation}
over all snapshots (time frames with active speech) $s = 1,\dots,S$, for each 10 s signal and all 120 source-microphone configurations (noting that the source position $\pbf[s]$ was static for each configuration). 
For the SRP-PHAT method, the TDOA estimation error was computed based on the TDOAs corresponding to the estimated source position. 
In addition, to evaluate the usability of the estimated TDOAs for SI-based source position estimation, we considered measures based on the position estimation error $\varepsilon[s]=||\hat{\pbf}[s] - \pbf[s]||_{2}$, i.e., the mean position estimation error over snapshots $\overline{\varepsilon} = \frac{1}{S} \sum_{n=1}^{S}  \; \varepsilon[s]$ 
and the accuracy $\mathrm{Acc}_{} = \frac{1}{S} \sum_{s=1}^{S} \mathcal{X} (\varepsilon[s]  \leq 10 \textrm{ cm})$, 
using the indicator function $\mathcal{X}(\cdot)$, which is equal to 1 if $\varepsilon[s] \,\leq\, 10\textrm{ cm}$ and 0 otherwise. 
\begin{table}[t!] 
\centering
\caption{Mean TDOA estimation error $\overline{\sigma}$, mean position estimation error $\overline{\varepsilon}$ and position estimation accuracy Acc$_{}$ for the SRP-PHAT method, three reference microphone-based methods (Ref-A, Ref-C, Ref-R), the baseline MST method, and the proposed MST+ method for three reverberation conditions.}
    \label{tab: results}
\newcommand\RowStretch{\rule[-5pt]{0pt}{15pt}}
\begin{tabular}{c|c|c|ccc|cc}
\toprule
\hline
\multirow{2}{*}{\hspace*{-1.5 mm}\rotatebox{90}{\textbf{Reverb.}}} & \textbf{Error} & \multicolumn{6}{c}{\RowStretch\textbf{TDOA Estimation Method}} \\ 
\cline{3-8}
\RowStretch & \textbf{Metric} & \begin{tabular}{@{}c@{}}\textbf{SRP-} \\ \textbf{PHAT}\end{tabular} & \textbf{Ref-A} & \textbf{Ref-C} & \textbf{Ref-R} & \textbf{MST} & \textbf{MST+} \\
\hline 
\multirow{3}{*}{\hspace*{-1.5 mm}\rotatebox{90}{{Low}}} 
   & $\overline{\sigma}$ [ms] & 0.18 & 0.49 & 0.16 & 0.04 & 0.04 & \textbf{0.01} \\
   & $\overline{\varepsilon}$ [cm] & 20.0 & 28.1 & 8.5 & 2.9 & 2.9 & \textbf{0.7} \\
   & Acc$_{}$ [\%]             & 69.5 & 81.1 & 93.9 & 98.2 & 98.3 & \textbf{99.8} \\ 
\hline
\multirow{3}{*}{\hspace*{-1.5 mm}\rotatebox{90}{{Med.}}}
   & $\overline{\sigma}$ [ms] & 0.28 & 1.34 & 0.54 & 0.14 & 0.12 & \textbf{0.08} \\
   & $\overline{\varepsilon}$ [cm] & 29.3 & 70.9 & 28.6 & 9.0 & 8.9 & \textbf{5.0} \\
   & Acc$_{}$ [\%]             & 63.0 & 54.3 & 77.1 & 92.9 & 92.8 & \textbf{95.8} \\
\hline
\multirow{3}{*}{\hspace*{-1 mm}\rotatebox{90}{{High}}}
   & $\overline{\sigma}$ [ms] & 0.44 & 1.77 & 0.80 & 0.44 & 0.40 & \textbf{0.28} \\
   & $\overline{\varepsilon}$ [cm] & 37.9 & 90.1 & 46.1 & 24.3 & 20.3 & \textbf{10.5} \\
   & Acc$_{}$ [\%]             & 59.6 & 44.2 & 62.7 & 82.2 & 85.0 & \textbf{92.4} \\
\hline
\bottomrule
\end{tabular}
\end{table}

For all considered TDOA estimation methods, \cref{tab: results} shows the TDOA and source position estimation errors for the three considered reverberation levels. 
As can be observed, the SRP-PHAT method results in a relatively poor source position estimation (and TDOA estimation) performance (e.g., $\overline{\sigma} = 0.28$ ms, $\overline{\varepsilon} =  29.3$ cm, and Acc$_{} = 63.0$ \% in medium reverberation). 
This is likely because the SRP-PHAT method has not been tested for such random source and microphone configurations, where it may not be ideal to weight the CPSD from each microphone pair equally within the SRP-PHAT functional. 
Considering the TDOA-based methods, randomly choosing the reference microphone (Ref-A method) also results in a relatively poor TDOA and source position estimation performance (e.g., $\overline{\sigma} = 1.34$ ms, $\overline{\varepsilon} =  70.9$ cm, and Acc$_{} = 54.3$ \% in medium reverberation). 
The results of the Ref-C method show that the TDOA and source position estimation performance can be easily improved by simply choosing the microphone closest to the centroid of the microphone array as the reference microphone (e.g., $\overline{\sigma} = 0.54$ ms, $\overline{\varepsilon} = 28.6$ cm, and Acc$_{} = 77.1$ \% in medium reverberation). 
The reliability-based Ref-R and MST methods outperform the SRP-PHAT, Ref-A and Ref-C methods, showing the importance of taking into account the reliability of the GCC-PHAT functions for determining the minimal set of microphone pairs. 
In low and medium reverberation conditions, the Ref-R and MST methods perform very similarly. 
However, in high reverberation the MST method performs slightly better than the Ref-R method ($\overline{\sigma} = 0.40$ ms, $\overline{\varepsilon} = 20.3$ cm, and Acc$_{} = 85.0$ \% compared to $\overline{\sigma} = 0.44$ ms, $\overline{\varepsilon} = 24.3$ cm, and Acc$_{} = 82.2$ \%), most likely because in the MST method no reference microphone needs to be chosen. 
The results in \cref{tab: results} clearly show that the proposed MST+ method outperforms all other considered methods in every reverberation condition (e.g., $\overline{\sigma} = 0.08$ ms, $\overline{\varepsilon} = 5.0$ cm, and Acc$_{} = 95.8$ \% in medium reverberation). 
This suggests that it is beneficial to average over multiple CPSDs from multiple microphone pairs to accurately estimate TDOAs and improve source position estimation performance, rather than estimating each TDOA based on a single CPSD. \vspace*{-3 mm}

\section{Conclusions}\vspace*{-0.2 mm}
In this paper, we have proposed an incremental method, improving the time-difference of arrival (TDOA) estimation based on an average over multiple cross-power spectral densities (CPSDs) from multiple microphone pairs, compared to reference microphone- and minimum spanning tree (MST)-based methods which rely on a single CPSD. 
This method re-estimates the TDOAs corresponding to the MST of generalized cross-correlation with phase transform (GCC-PHAT) function reliabilities, in multiple steps, beginning with the edges with highest GCC-PHAT reliabilities. 
In each step, we incorporate an additional CPSD, estimated indirectly via another microphone from a previous step. 
Including the indirectly estimated CPSD requires a phase-alignment, which we achieve using a phase shift based on re-estimated TDOAs from previous steps. 
Based on noisy and reverberant speech signals recorded in a laboratory with an array of spatially distributed microphones, we evaluated the performance of the different methods in terms of TDOA estimation error and source position estimation error, for three reverberation conditions. 
Experimental results for different source and microphone configurations demonstrate that the proposed method considerably improves the TDOA and source position estimation performance compared to existing reference microphone-, MST-based, and steered-response power-based methods.

\clearpage
\IEEEtriggeratref{16}
\bibliographystyle{IEEEtran}
\bibliography{ms}

\end{document}